\documentclass[superscriptaddress,
amsmath,amssymb,
aps,
prb,
twocolumn]{revtex4-2}

\usepackage{graphicx}
\usepackage{amsmath}
\usepackage{amssymb}
\usepackage{braket}
\usepackage{bbold}
\usepackage{hyperref}
\usepackage{subfigure}
\usepackage[usenames,dvipsnames]{color}
\usepackage{graphicx}
\usepackage{subfigure}
\usepackage[mathscr]{euscript}

\newcommand{\bcen}{\begin{center}}
\newcommand{\ecen}{\end{center}}
\newcommand{\btab}{\begin{tabular}}
\newcommand{\etab}{\end{tabular}}
\newcommand{\bdes}{\begin{description}}
\newcommand{\edes}{\end{description}}

\newcommand{\beq}{\begin{equation}}
\newcommand{\eeq}{\end{equation}}
\newcommand{\bea}{\begin{eqnarray}}
\newcommand{\eea}{\end{eqnarray}}

\newcommand{\bary}{\begin{array}}
\newcommand{\eary}{\end{array}}
\newcommand{\benum}{\begin{enumerate}}
\newcommand{\eenum}{\end{enumerate}}
\newcommand{\bitem}{\begin{itemize}}
\newcommand{\eitem}{\end{itemize}}

\newcommand{\Tr}{{\rm{Tr}}}

\newcommand{\Fig}[1]{Fig.~\ref{#1}}

\makeatletter

\newcommand{\Rmnum}[1]{\expandafter\@slowromancap\romannumeral #1@}
\makeatother

\newcommand{\tb}{\textcolor{black}}

\begin{document}

\title{Signatures of quantum phases in a dissipative system}

\author{Rohan Joshi}
\email{rajoshi2@illinois.edu}
\affiliation{Indian Institute of Technology Kanpur, Kalyanpur, Uttar Pradesh  208016, India}%
\affiliation{University of Illinois Urbana-Champaign, Urbana, Illinois 61801, United States}%
\author{Saikat Mondal}
\email{msaikat@iitk.ac.in}
\affiliation{Indian Institute of Technology Kanpur, Kalyanpur, Uttar Pradesh  208016, India}%
\author{Souvik Bandyopadhyay}
\email{sbandyop@bu.edu}
\affiliation{Indian Institute of Technology Kanpur, Kalyanpur, Uttar Pradesh  208016, India}
\affiliation{Department of Physics, Boston University, Boston, Massachusetts 02215, USA}%
\author{Sourav Bhattacharjee}
\email{sourav.bhattacharjee@icfo.eu}
\affiliation{Indian Institute of Technology Kanpur, Kalyanpur, Uttar Pradesh  208016, India}
\affiliation{ICFO - Institut de Ciències Fotòniques, The Barcelona Institute
of Science and Technology, Av. Carl Friedrich Gauss 3, 08860 Castelldefels (Barcelona), Spain}%
\author{Adhip Agarwala}
\email{adhip@iitk.ac.in}
\affiliation{Indian Institute of Technology Kanpur, Kalyanpur, Uttar Pradesh  208016, India}%

\begin{abstract}

Lindbladian formalism, as tuned to dissipative and open systems, has been all-pervasive to interpret non-equilibrium steady states of quantum many-body systems. We study the fate of free fermionic and superconducting phases in a dissipative one-dimensional Kitaev model  - where the bath acts both as a source and a sink of fermionic particles with different coupling rates. As a function of these two couplings, we investigate the steady state, its entanglement content, and its approach from varying initial states. Interestingly, we find that the steady state phase diagram retains decipherable signatures of ground state critical physics. We also show that early-time fidelity is a useful marker to find a subclass of phase transitions in such situations. Moreover, we show that the survival of critical signatures at late-times, strongly depend on the thermal nature of the steady state. This connection hints at a correspondence between quantum observables and classical magnetism in the steady state of such systems. Our work uncovers interesting connections between dissipative quantum many-body systems, thermalization of a classical spin and many-body quantum critical phenomena. 

\end{abstract}

\maketitle

\section{Introduction}

Quantum dissipative systems have emerged as one of the defining themes in recent quantum condensed matter research not only due to the fundamental theoretical questions it poses, but as well as a confluence of experiments in quantum information, cold-atomic systems, and material systems which are inherently out of equilibrium \cite{eisert2015quantum, langen2015ultracold, cazalilla2010focus, muller2012engineered, vasseur2016nonequilibrium, rotter2015review}. Further, the energetics of Floquet systems, questions in thermalization and equilibriation, and non-Hermitian Hamiltonians have, in parallel, provided alternate frameworks and thought points to pursue similar physics \cite{harper2020topology,oka2019floquet, mori2023floquet, ashida2020non, bergholtz2021exceptional, breuer2016colloquium}. Under a weak coupling and Markovian approximation where memory effects can be ignored - a particularly successful framework is developed by Gorini-Kossakowski-Sudarshan-Lindblad equation \cite{GoriniJMP2008, belavkin1969relaxation,lindblad1976generators}. Here instead of evolving a system wavefunction, a density matrix of the system itself is evolved with incoherent couplings to baths incorporated in the form of jump operators \cite{molmer1993monte}. While extensively pursued and discussed in the context of few-level systems, quantum optics setups, fairly recently, a huge body of work has been dedicated to their role in dissipative quantum many-body phases of matter \cite{eisert2015quantum, langen2015ultracold, cazalilla2010focus, muller2012engineered}. \tb{Moreover, recent works have investigated the non-equilibrium steady states of systems evolving under Lindbladian dynamics in the Zeno regime of large dissipation \cite{Presilla_PhysRevA_2017, Presilla_PhysRevA_2018, Presilla_PhysRevA_2020}.}

Free fermionic phases of matter, under dissipative settings where the jump operators are themselves linear can be exactly solved using the technique of third quantization~\cite{Prosen_2008, Prosen_PRL_2008, Prosen_2010}. A class of systems with both uniform and inhomogenous dissipators have been extensively studied to investigate steady states, low-lying excitations, the behavior of correlation functions, and entanglement signatures~\cite{bardyn2013topology}. Interestingly, it has been shown that in one-dimensional topological systems, edge modes decay exponentially fast in time for specific choices of jump operators~\cite{Carmele_PRB_2015} and multiple baths can lead to interference effects \cite{Abbruzzo_PRB_2021}. Topological invariants have also been defined for dissipative systems \cite{BudichPRB2015,ds21,Souvik_PRB_2020,mao_arxiv23}, and as well a complete symmetry classification of Lindbladians has been formulated \cite{LieuPRL2020}. In an interesting recent work, a universal master equation for such systems has also been devised \cite{NathanPRB2020}. While it is often assumed that Lindbladian evolution generically provides a mixed density matrix, it is interesting to ask the nature of bipartite entanglement entropy and mutual information as the system approaches a non-equilibrium steady state (NESS) \cite{Souvik_PRB_2020, CarolloPRB2022, Lepori_PRR_2022, Dutta_PRR_2021,Alba,Starchl_Sieberer}. Recently it was shown that in the Kitaev model, the system retains a finite mutual information even in the steady state which in general is of a mixed character \cite{MaityPRB2020, Moos_SP_2019}. Building on this, it is interesting to ask about the nature of the density matrix with regard to thermalization after sufficiently long times. For instance, does the steady state retain any decipherable signatures of the initial starting ground state?

This question has been partly addressed in some settings. Systems undergoing Linbladian dynamics are guaranteed to attain a thermal state with a well-defined temperature when the Kubo-Martin-Schwinger condition is satisfied \cite{breuer2002theory}. This leads to a complete loss of information of the initial state as the steady state is approached. However, this can be avoided in specially engineered settings. For example, there may exist certain dissipation-free subspaces of the Hilbert space which essentially evolve unitarily \cite{Diehl2011}. Similar phenomena can prevent any definition of a temperature in the asymptotic steady state of the system. Using simple toy models of quantum many-body systems, here, we show that it is indeed possible to extract ground-state critical information contained in the Hamiltonian even after long-time dissipative dynamics when the steady state is not thermal. The advantage of using such simple models is not just limited to analytical investigations but also to recent multidirectional experimental progress and ingenious proposals of realizing these systems using ultracold atomic lattices, quantum simulators, and quantum dots \cite{exp1,exp2,exp3,exp4,exp5,exp6,exp7,exp8,exp9,exp10,exp11,exp12,exp13,exp14,exp15}.

\begin{figure}
    \centering
    \includegraphics[width=1\linewidth]{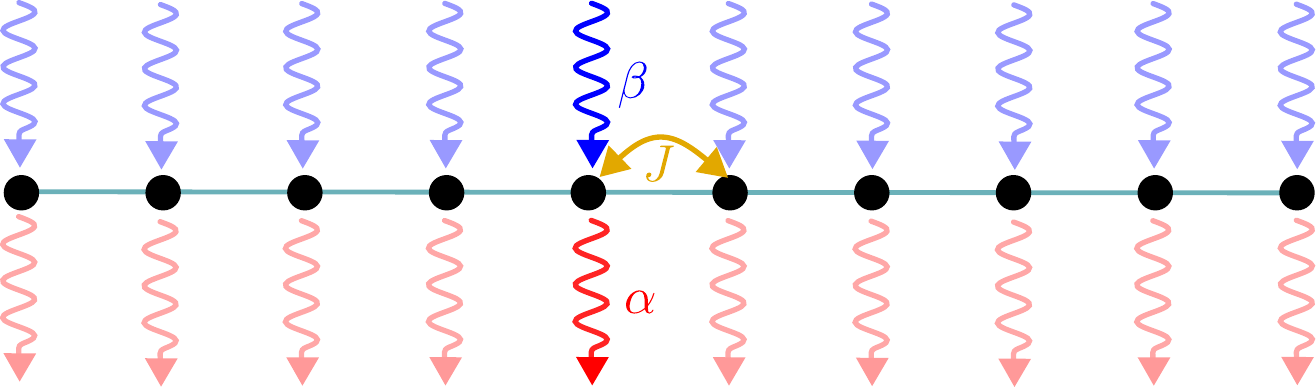}
    \caption{Schematic representation showing nearest-neighbour hopping $J$ and two bath couplings $\alpha$ and $\beta$ at each site (see Eq.~\eqref{eq_h}).}
    \label{fig:schematic}
\end{figure}

More specifically, we investigate the one-dimensional Kitaev model, with jump operators coupled to every site which can pump and annihilate fermions in the system with two different probability rates as shown in \Fig{fig:schematic}. We pose the question- can one engineer phase transitions within the steady states by tuning these parameters independently? Moreover what kind of entanglement transitions can the system generate? To this end, we uncover a host of interesting phenomena where steady-state physics is dominated by the bath parameters. Interestingly, in suitable parameter regimes, we map the steady state to thermal states which can effectively be mapped to a classical spin system in a magnetic field. We further study the mutual information and the behavior of correlations and provide a comprehensive picture of the dissipative physics in the Kitaev chain. We also show that apart from the steady states, even in early time dynamics, quantities such as fidelity can show signatures of ground-state phase transitions. Interestingly, host of this physics translates to higher dimensional systems where we show another example in terms of a two-dimensional Chern insulator model. We end with an outlook where we comment on the generalization of this physics to other systems and mention if there can be deviations particularly due to non-Markovian processes and interactions.

The paper is organized in the following way: in Section~\ref{sec:model} we introduce the model and the parameters. In Section~\ref{nessphy} we discuss steady-state physics, particularly focusing on the mapping to thermal density matrix, the various observables, the entanglement signatures, and fermionic correlations. In Section~\ref{earlytime}, we discuss the early time evolution of the system, in particular the survival probability or Loschmidt echo in the system. In Section~\ref{generalBHZ}, we generalize our results to the two-dimensional BHZ Chern insulator model and conclude with an outlook in section~\ref{sec:outlook}.

\section{Model}
\label{sec:model}

We study the Kitaev chain described by the Hamiltonian~\cite{Kitaev_2001}
\begin{equation}\label{eq_h}
    H = -\frac{1}{2}\sum_{n=1}^{L}(Jc_n^\dagger c_{n+1} - \Delta c_n c_{n+1} + h.c.) - \mu\sum_{n=1}^{L} \left(c_n^\dagger c_{n} - \frac{1}{2}\right),
\end{equation}
where $c_n$ and $c^\dagger_n$ are fermionic annihilation and creation operators at $n$-th site, $J$ is the hopping strength, $\Delta$ is the fermion pairing term and $\mu$ is the chemical potential. The system shows three distinct phases, of which two are topological (characterized by winding numbers $\pm 1$ respectively) and one is topologically trivial \cite{kitaev09,ds21}. The $\Delta = J$ line is identical to the one-dimensional transverse field quantum Ising model where $J$ maps to the ferromagnetic exchange and $\mu$ maps to the applied field along the z-direction.

Besides the unitary evolution due to $H$, we introduce a dissipative Lindbladian coupling of the form
\beq \label{lindbladian}
\frac{d \rho(t)}{dt} = -i[H, \rho(t)] + \alpha \sum_{n=1}^{L} \mathcal{D}[{\mathcal{L}}_{n}=c_{n}] + \beta \sum_{n=1}^{L} \mathcal{D}[{\mathcal{L}}_{n}=c^{\dagger}_{n}],
\eeq
where the `dissipator' $\mathcal{D}[{\mathcal{L}}_{n}]$ is given by,
\begin{equation}
    {\mathcal{D}}[{\mathcal{L}}_{n}] = {\mathcal{L}}_n\rho(t){\mathcal{L}}^\dagger_n -\frac{1}{2} \{ {\mathcal{L}}^\dagger_n {\mathcal{L}}_n, \rho(t) \}.
\end{equation}
Here, $\alpha$ and $\beta$ enter as parameters in the system. The effects of $\alpha$ and $\beta$ on the system are the primary focus of our analysis. We note that similar dissipative channels are expected in 1D cold-atom lattices interacting with  superconducting bath (see Ref.~\cite{Diehl2011}).

The specific form of the Lindbladian operators allows us to rewrite Eq.~\eqref{lindbladian} as,
\begin{equation}
\begin{split}
    \frac{d \rho(t)}{dt} = &-i[H, \rho(t)] + \sum_n (\alpha c_n \rho c^\dagger_n + \beta c^\dagger_n \rho c_n)\\
                      &-\sum_n \left(\frac{(\alpha - \beta)}{2} \{ c^\dagger_n c_n , \rho \} + \beta\rho\right).
\end{split}
\end{equation}

The quadratic nature of this evolution leads to decomposition into independent momentum sectors each spanned by the product basis states $| 0,0 \rangle$, $c^\dagger_k |0,0 \rangle = |k,0 \rangle$, $c^\dagger_{-k} |0,0 \rangle = |0,-k \rangle$, $c^\dagger_k c^\dagger_{-k} |0,0 \rangle = | k,-k \rangle$, which gives the momentum space evolution ($k>0$),
\begin{equation}\label{lindbladian_k}
\begin{split}
    \frac{d \rho^k(t)}{d t} = &-i [H_k, \rho^k(t)] +  (\alpha c_k \rho^k c^\dagger_k + \beta c^\dagger_k \rho^k c_k)\\
                        &+ (\alpha c_{-k} \rho^k c^\dagger_{-k} + \beta c^\dagger_{-k} \rho^k c_{-k})\\
                        &-  \left(\frac{(\alpha - \beta)}{2} \{ c^\dagger_k c_k , \rho^k \} + \beta\rho^k\right)\\
                        &-  \left(\frac{(\alpha - \beta)}{2} \{ c^\dagger_{-k} c_{-k}, \rho^k \} + \beta\rho^k\right),
\end{split}
\end{equation}
where the Hamiltonian $H_k$ is
\begin{equation}\label{Hamiltonian_Delta_k}
\begin{split}
    H_k &=  (\mu + J\cos(k))(c_{-k} c^\dagger_{-k} -  c^\dagger_k c_k)\\
        &- i\Delta\sin(k)(c^\dagger_{k}c^\dagger_{-k} - c_{-k} c_{k}).
\end{split}
\end{equation}

This decomposition into momentum sectors permits an exact solution for the steady state at $t=\infty$ which we now investigate.  The steady state is defined as  $\frac{d \rho_{ss}}{dt} = 0$ and is obtained as a Kronecker product over the steady states in the momentum sectors ($\rho_{\text{ss}} = \bigotimes_{k > 0}\rho^k_{ss}$). In the absence of pairing term $\Delta$, the system decouples into separate momentum ($k$) modes with the basis states $|0\rangle$ and $|k\rangle$ per momentum mode. We also recall that the critical points separating the ground state topological phases (in the absence of bath) of the Kitaev model correspond to the situation in which the bandgap of $H_k$ vanishes.
 
The Lindbladian in Eq.~\eqref{lindbladian} has time-reversal symmetry, particle-hole symmetry, and sublattice symmetry for all values of $\alpha$ and $\beta$ and thus it belongs to the BDI symmetry class~\cite{LieuPRL2020}. However, under the transformation $c_{n} \to (-1)^{n} c_{n}^{\dagger}$ and $\mu \to -\mu$, the dissipator $\mathcal{D}[\mathcal{L}_{n}=c_{n}]$ transforms to $\mathcal{D}[\mathcal{L}_{n}=c_{n}^{\dagger}]$ and vice-versa keeping the Hamiltonian $H$ invariant. These symmetries play a useful role to understand the results as we discuss in the following sections. 

\section{Non-Equilibrium Steady States}   
\label{nessphy}

The steady-state phase diagram has the following distinct characteristics depending on whether the pairing term is zero/non-zero and relative strengths of $\alpha, \beta$.  We find that in the absence of the pairing term, irrespective of the $\alpha, \beta$ couplings, the system always stabilizes to a `thermal state' characterized by a diagonal (in the product basis) density matrix which can be assigned an effective temperature and a Zeeman energy scale as we discuss in this section. Also, if one of the bath couplings is identically zero, the system evolves to a pure direct product steady state. On the other hand, in the presence of the pairing term the physics is subtly different. The steady state in this situation cannot be directly mapped to any effective thermal state (for $\alpha \neq \beta$) and also retains a finite non-diagonal contribution in the product basis. The physics of this non-diagonal piece reflects a depleted superconductor. The results are summarized in Table.\ref{tab:ssans}. We first discuss the physics in the absence of a pairing term and then introduce the role of pairing term.

\begin{table}[h!]
    \begin{tabular}{|c|c|c|} \hline
         & $\Delta = 0$  & $\Delta \neq 0$ \\ \hline \hline
        $\alpha \neq \beta \neq 0$ & Thermal  & Non-thermal  \\ 
        & mixed state ($0<\zeta<\infty$) & mixed State \\
        \hline
       $\alpha =0, \beta \neq 0$   & Thermal  & Non-thermal  \\ 
        \text{or vice versa}& pure state ($\zeta=0$) & mixed state \\
       
       \hline
       $\alpha = \beta \neq 0$   & Thermal   & Thermal  \\ 
        & mixed state ($\zeta=\infty$) & mixed state ($\zeta=\infty$) \\
       \hline
    \end{tabular}
    \caption{Nature of the $t=\infty$ steady state depending on the bath parameters and pairing term. The steady state density matrix for $\Delta = 0$ is given by Eq.~\eqref{eq_rho_thermal}. For a general $\Delta \neq 0$ the steady state is given in Appendix~\ref{app_sskt}.}
    \label{tab:ssans}
\end{table}

\subsection{No pairing~~$\Delta=0$~: U(1) fermions}
We first consider the case with $\Delta = 0$ such that we have a simple tight-binding Hamiltonian $H$ having a hopping amplitude $J$ and chemical potential $\mu$ with a $U(1)$ number conservation for the fermions,
\beq
H = -\frac{J}{2}\sum_{n=1}^{L} (c^\dagger_n c_{n+1} + \text{ h.c.}) - \mu\sum_{n=1}^{L} \left(c^\dagger_n c_n - \frac{1}{2}\right).
\eeq

This gives us the momentum space evolution ($-\pi<k<\pi$),
\begin{equation}\label{eq_lind_U1}
\begin{split}
    \frac{d\rho^k(t)}{dt} = &-i [H_k , \rho^k(t)] +  (\alpha c_k \rho^k c^\dagger_k + \beta c^\dagger_k \rho^k c_k)\\
                        &-  \left(\frac{(\alpha - \beta)}{2} \{ c^\dagger_k c_k , \rho^k \} + \beta\rho^k\right)\\
\end{split}
\end{equation}
where now the Hamiltonian $H_k$ is,
\beq
    H_k =  -(\mu + J\cos(k))c^\dagger_k c_k.
\eeq

 Under the action of the bath operators, the steady state ($\frac{d\rho_{ss}}{dt} = 0$) is obtained as a Kronecker product over the momentum sectors, $\rho_{ss}=\bigotimes_k\rho^k_{ss}$, where  $\rho^k_{ss}$ is diagonal in the basis spanned by the basis states $|0\rangle$, $|k\rangle=c^\dagger_k|0\rangle$ for each $k$,
\begin{center}
$\rho_{\text{ss}}(k) =
 \begin{pmatrix}
 \frac{\alpha}{\alpha +\beta} & 0 \\
 0 & \frac{\beta}{\alpha +\beta} \\
 \end{pmatrix} .$  
\end{center}
 There are two interesting things to notice: (i) it is independent of $k$ and also the corresponding energy gap $\epsilon_k$ - thus {\it any} translationally invariant quadratic fermionic Hamiltonian with local bath and source-sink jump operators meets the same fate. (ii) If one of the bath rates i.e.~$\alpha$ or $\beta$ is made zero, the system reaches a pure state. Interestingly this pure state also gets annihilated by the corresponding jump operators and commutes with the Hamiltonian; hence it is a dark state \cite{bardyn2013topology}. Thus the observables in the infinite time steady state would be unable to detect ground state free fermionic phase transitions.
 
The above results can be understood as follows. A careful look at Eq.~\eqref{eq_lind_U1} reveals that the dissipative evolution for the $U(1)$ charge in the momentum space corresponds to the case in which each of the momentum modes can be considered to be a spin-$1/2$ coupled to a Zeeman field and a thermal bath. Here, $\alpha$ and $\beta$ correspond to the Boltzmann weights of the two respective spin states such that their asymmetric component ($\beta- \alpha$) is proportional to the Zeeman field strength. In the absence of any field, the density matrix reduces to an equal probability distribution. We shall make some of these correspondences more quantitative below. Thus we identify $\alpha/\beta = \exp\left(- 1/\zeta   \right)$ where $\zeta$ is an equivalent temperature scale with field strength set to unity. Each of the momentum modes in this case is therefore driven to the thermal steady state
\beq \label{eq_rho_thermal}
\rho_{\text{thermal}}(k) =
\begin{pmatrix}
\frac{e^{-\frac{1}{2 \zeta}}}{e^{-\frac{1}{2 \zeta}} +e^{\frac{1}{2 \zeta}}} & 0 \\
0 & \frac{e^{\frac{1}{2 \zeta}}}{e^{-\frac{1}{2 \zeta}} +e^{\frac{1}{2 \zeta}}} \\
\end{pmatrix}=\rho_{\text{ss}}(k).
\eeq

It is thus clear that for $\alpha=\beta$, we have $\zeta=\infty$, and the system is driven to the infinite temperature thermal steady state. On the contrary, $\alpha (\beta)=0$ corresponds to the zero temperature thermal steady state, which is consequently a pure state.   

Now, the expectation value of the number density $\hat{n}= \frac{1}{L} \sum_{j=1}^{L} c_{j}^{\dagger} c_{j}$ in the steady state, irrespective of the initial chemical potential, is given by 
\beq
\Tr (\rho_{ss} \hat{n} ) = \frac{\beta}{\alpha +\beta},
\eeq
which is thus just given by rates of sink and source baths. For an effective spin $\frac{1}{2}$, this can be interpreted as the magnetization of the system in the $z$ direction as 
\beq \label{magmap}
m_z = \frac{1}{L} \sum_{j=1}^{L} \Tr ( \rho_{ss} \sigma_{j}^{z}) =\Tr \left[\rho_{ss} (2 \hat{n} - 1)\right] = \frac{\beta-\alpha}{\beta+\alpha},
\eeq
or the effective magnetization $m_{z}$ is proportional to the anisotropy $h \equiv (\beta-\alpha)$ between the bath strengths. Here, the spin $\frac{1}{2}$ is mapped to the spinless fermion, such that   $\sigma_{j}^{z}=(2c_{j}^{\dagger} c_{j}-1)$ which follows from Jordan-Wigner transformation~\cite{lieb_1961}. The corresponding ``susceptibility" can be interpreted as 
\beq \label{eq_chi_m}
\chi_m =\frac{\partial m_z}{\partial h}= \frac{\partial m_z}{\partial (\beta - \alpha)} \sim \frac{1}{(\alpha + \beta)}.
\eeq
Eq.~\eqref{magmap} and Eq.~\eqref{eq_chi_m} allow us to define an ``effective temperature scale" $T_{\rm{eff}} \equiv \alpha+\beta$ and an ``effective magnetic field" $h \equiv \beta-\alpha$, mirroring the Curie's law.
 {\tb{It is noteworthy that although $\alpha$ and $\beta$ are not the actual temperature scales, they have dimensions of energy similar to temperature and magnetic field in the natural units where the Boltzmann constant and magnetic moment are considered unity. Thus, $(\alpha+\beta)$ and $(\beta-\alpha)$ can be interpreted as effective temperature $T_{\rm{eff}}$ and effective magnetic field $h$, leading to the analogy of spin $\frac{1}{2}$ in presence of a magnetic field and temperature. We also note that $T_{\rm{eff}}$ is different from $\zeta$ defined near Eq.~\eqref{eq_rho_thermal} where the magnetic field is set to unity.}}

\begin{figure}
    \centering
    \includegraphics[width=1\linewidth]{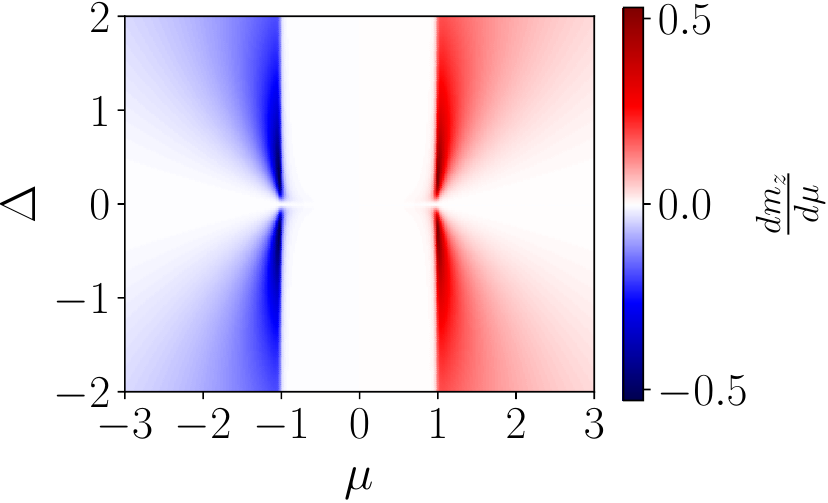}
    \caption{$\frac{d m_z}{d\mu}$ (derivative of magnetization $m_{z}$ with respect to chemical potential $\mu$) in the steady state as a function of $\mu$ and $\Delta$ for $\alpha \neq \beta$. While the $\Delta=0$ transition line cannot be captured, at a finite pairing ($\Delta \neq 0$) the critical lines at $\mu=\pm 1$ are still identifiable in the steady state. Plot is generated for $J = 1$, $\alpha = 0.015$, $\beta = 0.035$, system size $L = 500$. }
    \label{fig:fig1}
\end{figure}

\subsection{Finite pairing~~$\Delta \neq 0$~: Superconducting Systems}

\begin{figure*}
    \centering
    \centering
\subfigure[]{
\includegraphics[width=0.45\linewidth]{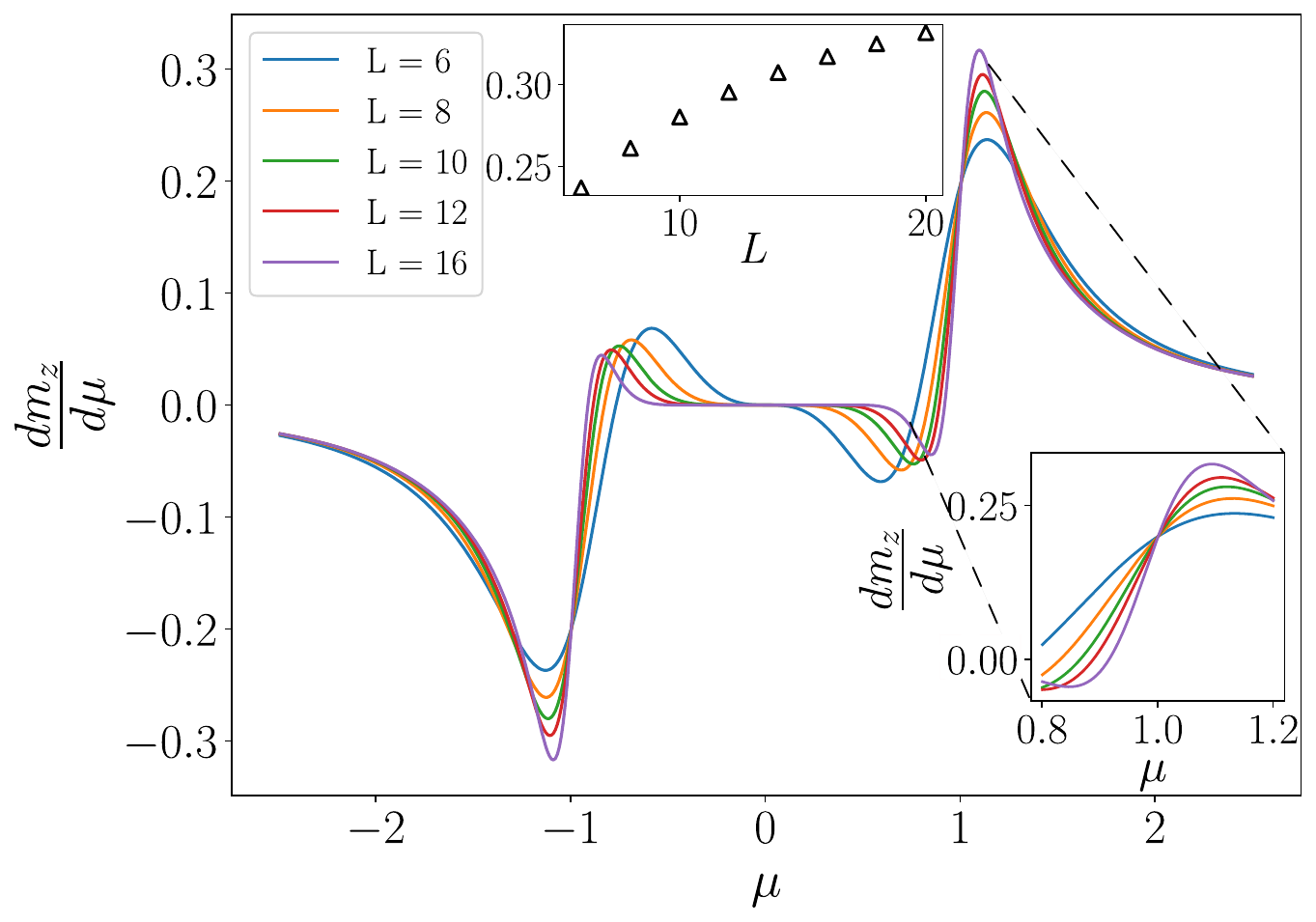}}
\centering
\hspace{5mm}
\subfigure[]{
\includegraphics[width=0.45\linewidth]{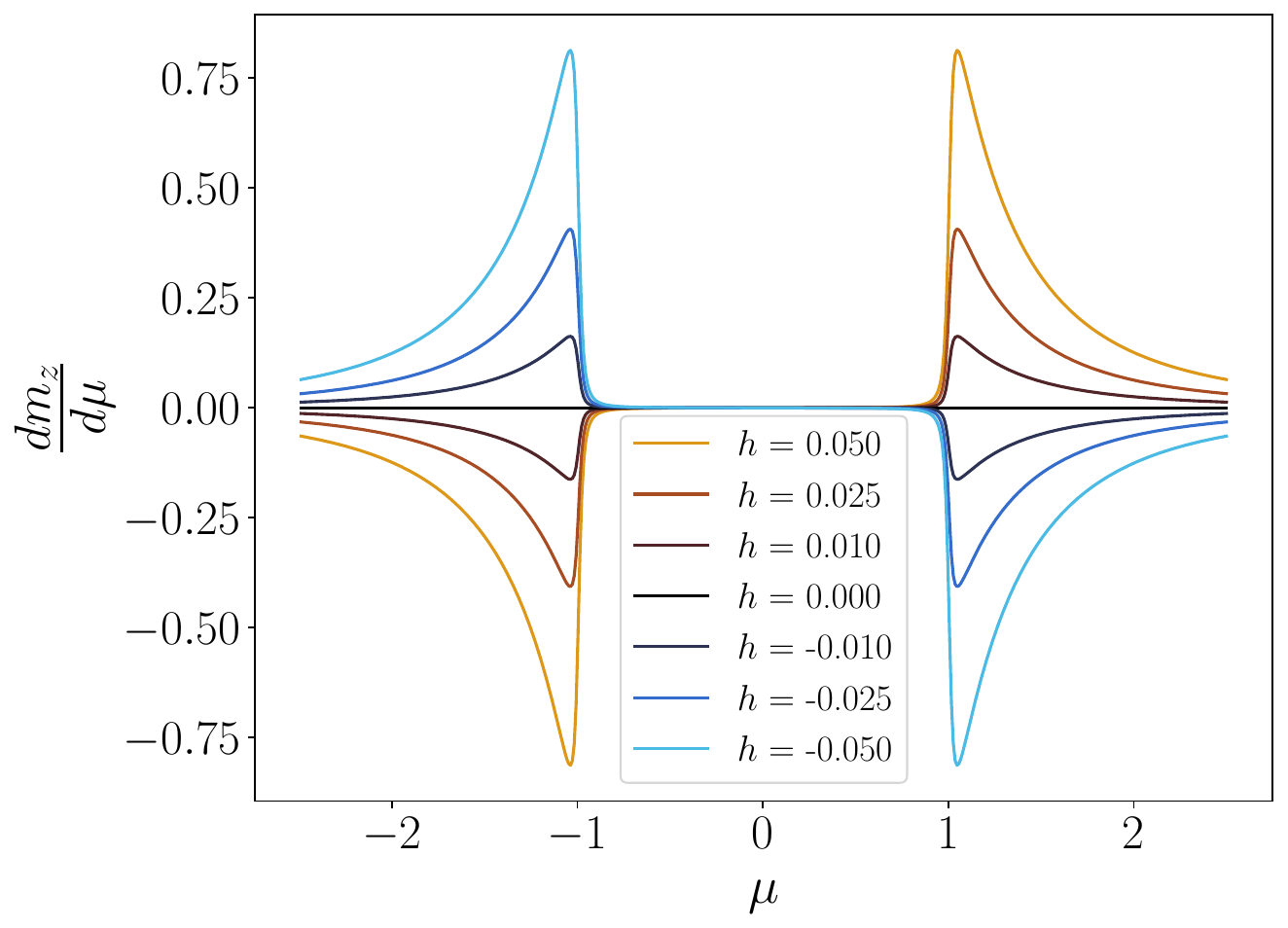}}
    \caption{(a) System size dependence of $\frac{dm_{z}}{d\mu}$ for $\alpha = 0.015, \beta = 0.035$. Inset (top): System size scaling of $\left|\frac{d m_{z}}{d\mu}\right|$ near $\mu = \pm 1$ critical points. (b) Coupling asymmetry of $\frac{d m_{z}}{d\mu}$ about the $h = \beta - \alpha = 0$ case, with $L = 500$ and $T_{\text{eff}} = (\alpha+\beta) = 0.05$. For all the plots, $J = \Delta = 1$.} 
    \label{fig2}
\end{figure*}

We now consider the time evolution described by Eq.~\eqref{lindbladian_k}, with the Hamiltonian $H_k$ as per Eq.~\eqref{Hamiltonian_Delta_k} (containing a pairing term $\Delta$) for the mode with momentum $k>0$. In this situation, it is instructive to observe that the dissipative evolution corresponds to a situation where the thermal baths can induce excitations only between certain states of the system. To elaborate, the decoupled modes for $\Delta\neq 0$ comprise of four-level systems with basis states $|0,0 \rangle, |k,0 \rangle, |0,-k \rangle, |k,-k \rangle$. However, the Lindblad operators are linear, which implies that these can only lead to single-particle excitations in the system. In particular, we note that the Lindblad operators can not lead to the excitations between $|0,0 \rangle \leftrightarrow |k,-k \rangle$, and $|k,0 \rangle \leftrightarrow |0,-k \rangle $. It therefore follows that the system cannot be expected to attain a thermal steady state.

Despite the non-trivial nature of the evolution, it is possible to gain some insight into the behavior of the steady state. We first note that $\alpha$ and $\beta$ control the rate at which the dissipation tries to drive the system to the vacuum state $|0,0 \rangle$ and fully occupied state $|k,-k \rangle$, respectively. In parallel, the Hamiltonian $H$ driving the unitary part of the evolution generates superconducting excitations (for $\Delta\neq 0$), thus entangling the $k$ and $-k$ modes. For $\alpha\neq 0, \beta=0$, the combined effect of these processes results in a steady state with populations  $ \langle k,0 | \rho_{ss} | k,0 \rangle=\langle 0,-k |\rho_{ss} | 0,-k \rangle= \langle k,-k | \rho_{ss} | k,-k \rangle < \langle 0,0 | \rho_{ss}| 0,0 \rangle$ and non-zero off-diagonal elements $\langle 0,0 | \rho_{ss}| k,-k \rangle= \langle k,-k | \rho_{ss}| 0,0 \rangle^*\neq 0$. Similarly, for $\alpha= 0, \beta\neq 0$, the steady state populations satisfy $\langle k,0 |\rho_{ss}|k,0 \rangle=\langle 0,-k | \rho_{ss}|0,-k \rangle=\langle 0,0 |\rho_{ss}|0,0 \rangle < \langle k,-k | \rho_{ss}|k,-k \rangle$, and have a non-vanishing off-diagonal element with the opposite sign (but otherwise identical) as that in the $\alpha\neq 0, \beta=0$ case. Finally, when the dissipation rates are equal, i.e., $\alpha=\beta\neq 0$, the diagonal elements are thus identical. In addition, the off-diagonal elements $\langle 0,0|\rho_{ss}|k,-k \rangle$ and $\langle k,-k | \rho_{ss}|0,0 \rangle$ also vanish as their steady-state values are proportional to the imbalance between $\langle k,-k |\rho_{ss}| k,-k \rangle$ and $\langle 0,0 | \rho_{ss}|0,0 \rangle$. The steady state in this case is therefore maximally mixed with $\zeta=\infty$. The matrix elements for the steady state density matrix for arbitrary $\alpha$ and $\beta$ are provided in Appendix~\ref{app_sskt}.

In general, when the superconducting pairing term is non-zero, the system retains a finite superconducting order parameter in the steady state. Interestingly the average number density in the system is modified with a correction term as,

\beq 
\Tr ( \rho_{ss} \hat{n} ) = \frac{\beta + f(\Delta)}{\alpha +\beta},
\eeq

where 
\beq \label{eq_fd}
f(\Delta) = \int_0^\pi \frac{dk}{2\pi} \frac{4(\alpha-\beta)\Delta^{2} \sin^{2}(k)}{\left( (\alpha+\beta)^2 + 4 \Delta^2 \sin^{2}(k) + 4 (\mu + J\cos(k))^{2} \right)}.
\eeq

The integral in Eq.~\eqref{eq_fd} can be rewritten as
\begin{equation}
    (\alpha-\beta)^{-1}f(\Delta) = \frac{1}{2\pi}\int_0^\pi dk\frac{4\Delta^{2} \sin^{2}(k)}{ T_{\rm{eff}}^2 + \epsilon_k^2 },
\end{equation}
where $\epsilon_k=\sqrt{4 \Delta^2 \sin^{2}(k) + 4 (\mu + J\cos(k))^{2}}$ is the energy gap of the Hamiltonian and $T_{\text{eff}}=\alpha+\beta$ is the average bath coupling energy scale. This effective energy scale regularizes any singularities in the integral at the critical point, smoothing out any non-analytic behavior due to the Hamiltonian gap closing. However, even for finite $T_{\rm{eff}}$, we see observable footprints of the critical point in the dissipative steady state, which despite their non-singular nature accurately detect the critical point.

One can thus conclude that the number density and related observables, such as magnetization evaluated in the steady state can signal gap closing points of the closed system when $\Delta \neq 0$ and $\alpha \neq \beta$. Relating to the analogy of the magnetization as mentioned in Eq.~\eqref{magmap}, we thus expect that $\frac{dm_z}{d\mu}$ when evaluated in the steady state would still carry signatures of the equilibrium phase diagram. We also find that the critical point at $\Delta=0$ remains undetected (see Fig.~\ref{fig:fig1}). {\tb{This can be explained as follows: for $\alpha \neq \beta$ and $\Delta \neq 0$, the system reaches a non-thermal steady state (see Table~\ref{tab:ssans}) and the steady state can still contain information about the initial ground state (of the closed system without any bath). When $\Delta=0$, the steady state is a thermal state and thus the critical point $\Delta=0$ remains undetected in the steady state.}}
In \Fig{fig2}, we plot $\frac{dm_z}{d\mu}$ as a function of $\mu$ for different system sizes and different values of $h=\beta-\alpha$. It is important to note that both $h$ and $\Delta$ need to be finite to decipher the ground state critical points. We also remark that unlike in the ground state, the observables in the NESS do not diverge at the critical points with increasing system size. This can be understood from Eq.~\eqref{eq_fd} where the coupling energy $\alpha + \beta$ broadens the critical region, thus effectively smearing out the critical singularity. 

\begin{figure*}
\includegraphics[width=1\linewidth]{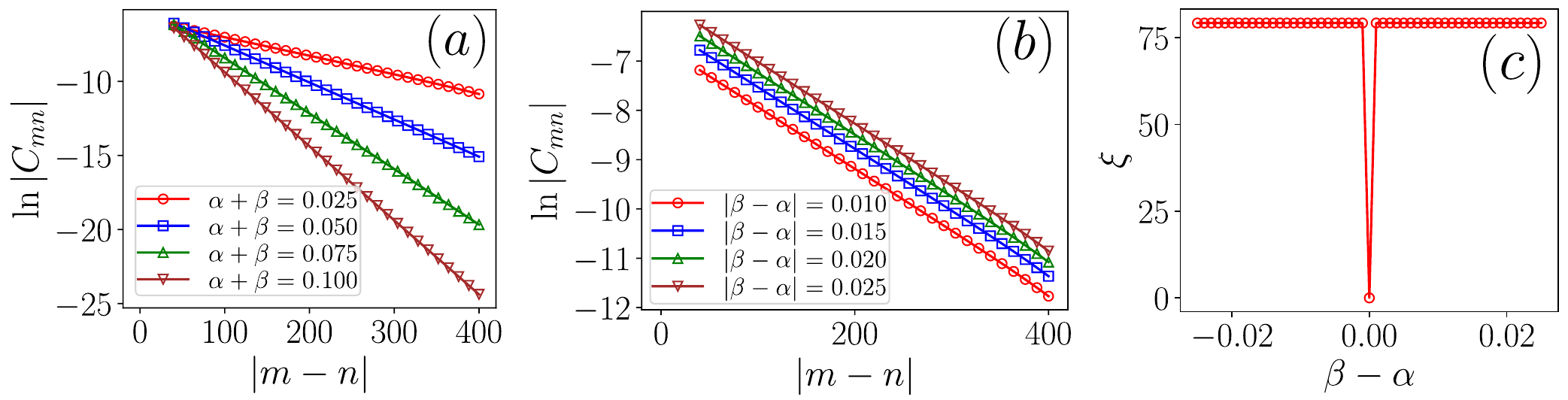}
        \caption{(a) $\ln|C_{mn}|$ (where the two-point correlation function $C_{mn}$ in the steady state is defined by Eq.~\eqref{C_mn_eqn}) in a dissipative Kitaev chain as a function of $|m-n|$. Here, $|C_{mn}|$ decays exponentially with $|m-n|$ and the decay rate depends on $\alpha + \beta$ when $\alpha \neq \beta$ and $\Delta \neq 0$. \tb{(b) $\ln|C_{mn}|$ in the steady state as a function of $|m-n|$ for fixed $(\alpha+\beta)=0.025$. (c) Correlation length $\xi$ as a function of $(\beta-\alpha)$ when $\alpha+\beta=0.025$. In this plot, $\xi$ is obtained from the slope of the plot of $\ln|C_{mn}|$ with $|m-n|$ (i.e. $\xi=-1/{\rm{slope}}$). In all the plots, we have chosen $\mu = J = \Delta = 1$, $L = 1000$.}}\label{C_cut} 
\end{figure*}
\subsection{Entanglement and Correlations}
The ground state of the one-dimensional Kitaev model has distinctive signatures in the correlations and entanglement in each of the phases. It is therefore natural to investigate their fate in the NESS. We study the two-point correlation function in the steady state defined as,
\beq \label{C_mn_eqn}
 C_{mn} = \langle c^\dagger_m c_n \rangle = \Tr(\rho_{\text{ss}} c^\dagger_m c_n).
\eeq
For $\Delta=0$, it can be shown that,
\beq 
\langle c^\dagger_i c_{i+r} \rangle = \delta(r) \frac{\beta}{\alpha+\beta},
\eeq
and similarly for $\alpha=\beta$, irrespective of value of $\Delta$,  
\beq \label{eq_corr_equal}
\langle c^\dagger_i c_{i+r} \rangle = \delta(r) \frac{1}{2},
\eeq
representing a highly short-range correlated state. For $\Delta \neq 0$, $\alpha \neq \beta$, near a critical point $\langle c^\dagger_i c_{i+r} \rangle  \sim \exp(-r/\xi)$  where $\xi \propto \frac{2 \Delta}{\alpha+\beta}$.  Thus the exponential decay is decided by the correlation length which is inversely proportional to the dissipation scale $\alpha +\beta$ (see Fig.~\ref{C_cut}(a)).
\tb{We also find that for a fixed $(\alpha+\beta)$, the correlation length $\xi$ (where $\xi$ is $-1/\rm{slope}$ of the plot of $\ln|C_{mn}|$ with $|m-n|$) is independent of $|\beta-\alpha|$ when $\alpha \neq \beta$ and $\Delta \neq 0$ (see Fig.~\ref{C_cut}(b,c)). When $\alpha=\beta$, the correlation length is zero ($\xi=0$) due to the behavior of correlation function as shown in Eq.~\eqref{eq_corr_equal}.}

A physical state with exponentially decaying correlations would in general imply an area-law bipartite entanglement entropy {(in the thermodynamic limit)} - however, this holds true only for a strictly pure state. In general, for a mixed state, given the complete system (excluding the bath) itself has a finite entanglement entropy, a more well-posed object of interest to analyze entanglement between two subregions A and B {in a mixed state} is the mutual information (MI),
\beq\label{MI_eqn}
    I(A,B) = S(A) + S(B) - S(A \cup B),
\eeq
where $S(A)$, $S(B)$ and $S(A \cup B)$ are the von Neumann entropies of the subsystems $A$, $B$ and the full system $A \cup B$. We note that the quantity $S(A \cup B)$ has a volume law scaling with the system size. This is not unexpected since each site of the system is uniformly coupled to the bath; thus the number of particle exchange channels, generating mixedness of the density matrix, also grows linearly with the system size~\cite{MaityPRB2020}. The mutual information is evaluated using the correlator matrix method generalized to mixed density matrices~\cite{Peschel_2003}(also see additional details in~\cite{MaityPRB2020}).

\begin{figure}
        \includegraphics[width=1.0\columnwidth]{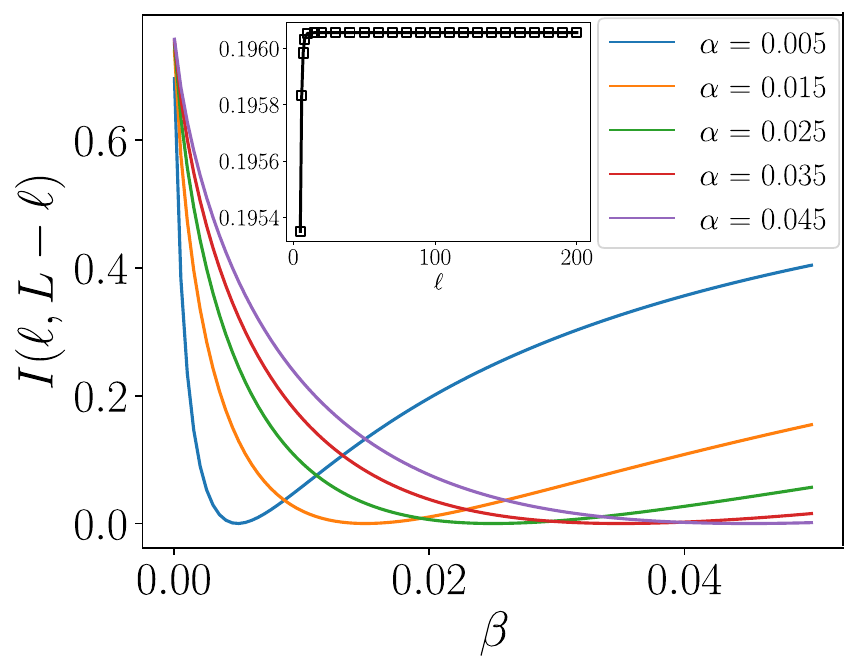}
        \caption{\label{MI} The mutual information $I(\ell; L-\ell)$ as defined in Eq.~\eqref{MI_eqn} shows an area law behaviour for all $\alpha$, $\beta$, and vanishes for $\alpha = \beta$ (plots are for $\ell = 20$). Inset: $I(\ell; L-\ell)$ variation with subsystem size $\ell$, for $\alpha = 0.01, \beta = 0.04$. Here $J = \Delta = \mu = 1$, $L = 500$.}
\end{figure}

We have already seen that when $\alpha=\beta$, the system approaches an infinite temperature steady state, irrespective of the magnitude of the pairing term $\Delta$. This leads to volume law entanglement entropy for any real space partition and mutual information goes identically to $\rightarrow zero$. At any general $\alpha \neq \beta$ the system still consists of a volume law entanglement entropy {(due to extensive entanglement between the system and the bath)} and area law MI given exponentially decaying fermionic correlations. The results are shown in \Fig{MI}.

With this, we complete the discussion of the steady state physics. Our results illustrate that the interplay of the pairing term as well as finite dissipative source and sink rates provide decipherable signatures of the ground state phases. In the next section, we discuss if early time evolution can also provide signatures of the ground state phase transitions.

\section{Early time Behavior}
\label{earlytime}

In recent studies \cite{adas_scirep_15,sthitadhi_adas_prb_17,ad21,apd23,ceren_prb_23}, long-time dynamics of quantum information measures such as the Loschmidt echo have been used to detect signatures of quantum critical physics in thermalizing systems. While these studies focus on the microcanonical picture of thermalization and the extraction of critical exponents in the steady state, in this paper we have seen that it is also possible to detect the critical point in the dissipative steady state of an open system. We now discuss how the early time dynamics of information measures such as the survival probability capture ground state critical physics, soon after the system is coupled to both the source and sink baths. 

\subsection{Survival probability}
\begin{figure*}
    \centering
\includegraphics[width=1.0\linewidth]{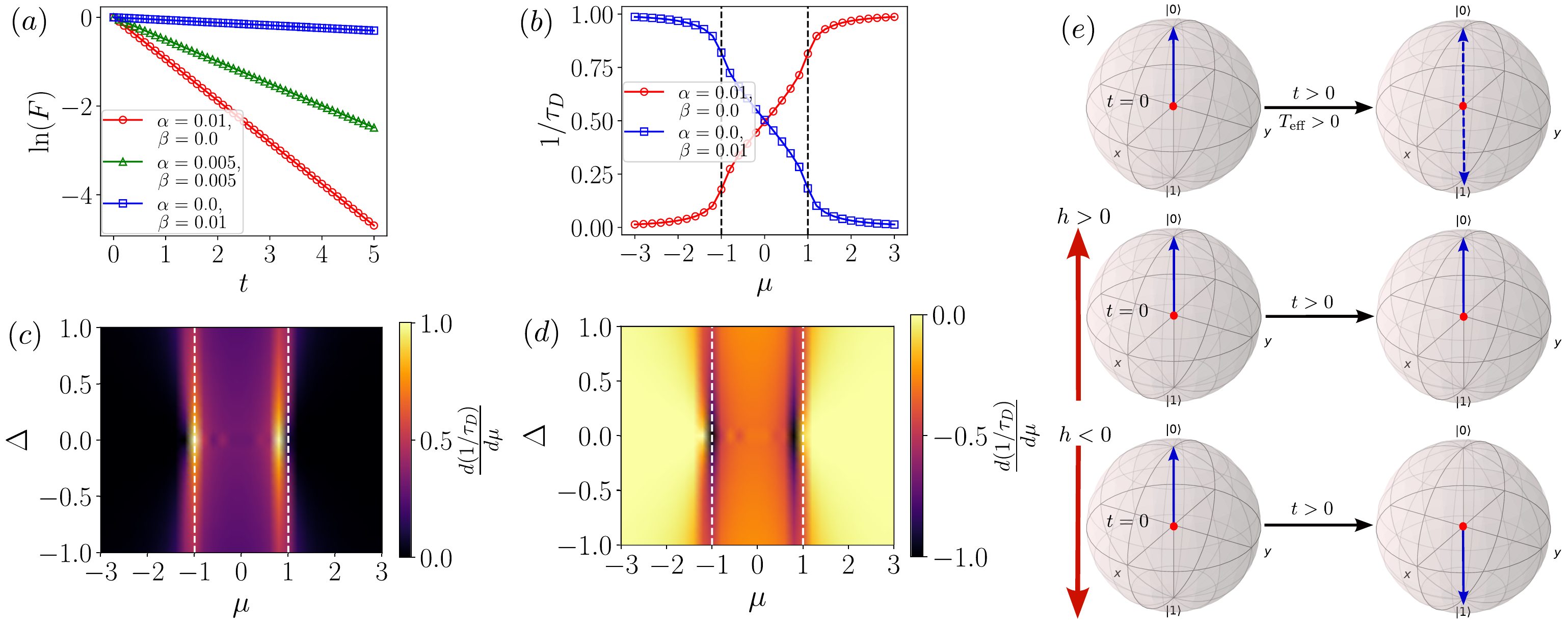}
     \caption{(a) $\ln(F)$ as a function of time $t$ for $\mu=1.5$ (where  $F(t)$ is the survival probability, as defined in Eq.~\eqref{eq_sp}), (b) $1/\tau_{D}$ (where $F(t) \sim \exp(-t/\tau_{D})$) as a function of $\mu$, (c) $\frac{d(1/\tau_{D})}{d\mu}$ as a function of $\mu$ and $\Delta$ for fermion-decaying bath ($\alpha=0.01$, $\beta=0$), (d) $\frac{d(1/\tau_{D})}{d\mu}$ as a function of $\mu$ and $\Delta$ for fermion-pumping bath ($\alpha=0$, $\beta=0.01$), (e) schematic diagram of a spin at time $t=0$ and $t>0$ in presence of an external magnetic field $h$ and temperature $T_{\text{eff}}$. In (a,b,c,d), we have chosen periodic boundary conditions of the Kitaev chain with $J=1.0$ and $L=100$. In (a,b), we have taken $\Delta=1.0$. In (b,c,d), $1/\tau_{D}$ is obtained numerically from the slope of the plot of $\ln(F)$ with $t$, where $t \in [0,5]$. In (b,c,d), the dotted lines at $\mu=1$ and $\mu=-1$ are the critical lines.}
\label{fig:survprob_plot}
\end{figure*}

Let us consider the Kitaev chain in the ground state $\ket{\psi (0)}$ of the Hamiltonian $H$ (see Eq.~\eqref{eq_h}) and the corresponding density matrix is given by $\rho(0)=|\psi (0) \rangle \langle \psi (0)|$. At $t=0$, the chain gets coupled to the bath, such that the dynamics of the reduced system at time $t>0$ is determined by the Lindbladian equation (see Eq.~\eqref{lindbladian}). However, it is noteworthy that the chain is not subjected to the quench of any parameter of the Hamiltonian.
 The survival probability~\cite{delcampo17} of the initial state at $t>0$ is then defined as
\begin{equation}\label{eq_sp}
F(t) = \frac{\Tr (\rho (t) \rho(0))}{\Tr(\rho^{2}(0))},
\end{equation}
where $\rho (t)$ is the reduced density matrix at time $t$.
At early time $t$, we find that $F(t) \sim \exp(-t/\tau_{D})$, as evident from Fig.~\ref{fig:survprob_plot}(a). At small time $t\rightarrow 0$, the survival probability $F(t)$  reduces to the following form~\cite{delcampo17}:
\begin{equation}\label{eq_sp_os}
F(t) \approx 1 - \frac{t}{\tau_{D}} + \mathcal{O} (t^{2}).
\end{equation}
Here the time-scale $\tau_{D}$ can be computed using the following equation (see Appendix~\ref{app_tau}):
\begin{equation}\label{eq_tau_kitaev}
\frac{1}{\tau_{D}}= \sum_{n=1}^{L} \Big( \alpha \Tr (c^{\dagger}_{n} c_{n} \rho(0)) + \beta \Tr ( c_{n} c^{\dagger}_{n} \rho(0)) \Big).
\end{equation}
When $\mu>0$, $F(t)$ decays faster for the fermion-decaying bath ($\alpha \neq 0$, $\beta=0$) than for fermion-pumping bath ($\alpha=0$, $\beta \neq 0$). Interestingly, the behavior of $F(t)$ becomes opposite when $\mu<0$ (see Fig.~\ref{fig:survprob_plot}(b)), which can be explained as follows: under the transformation $c_{n} \to (-1)^{n} c_{n}^{\dagger}$ (for all $n$) and $\mu \to -\mu$, the Hamiltonian $H$ remains invariant, but the dissipator $\mathcal{D}[{\mathcal{L}}_{n}=c_{n}]$ transforms to $\mathcal{D}[{\mathcal{L}}_{n}=c^{\dagger}_{n}]$. This suggests that $\tau_{1}(\mu)=\tau_{2}(-\mu)$, where $\tau_{1}$ and $\tau_{2}$ are the time-scales that determine the decay of $F(t)$ and are associated with fermion-decaying bath ($\alpha \neq 0$, $\beta=0$) and fermion-pumping bath ($\alpha=0$, $\beta \neq 0$) respectively. Further, for $\alpha=\beta \neq 0$, the decay-rate of $F(t)$ is independent of $\mu$.

The time-scales $\tau_{D}$ in presence of fermion-decaying bath ($\alpha \neq 0$, $\beta=0$) and fermion-pumping bath ($\alpha=0$, $\beta \neq 0$)
are found to be significantly different in the regions $\mu>1$, $|\mu|<1$ and $\mu<-1$, where we set $J=1$. The derivative of $1/\tau_{D}$ with respect to $\mu$ (i.e. $\frac{d(1/\tau_{D})}{d\mu}$) shows sharp changes at $\mu=\pm 1$ for both fermion-decaying bath ($\alpha \neq 0$, $\beta=0$) and fermion-pumping bath ($\alpha=0$, $\beta \neq 0$), as shown in Fig.~\ref{fig:survprob_plot}(c) and (d). Thus, the detection of ground-state phases and the critical lines located at $\mu =\pm 1$ is possible using $\tau_{D}$. However, the critical line at $\Delta=0$ remains undetected, as there is no significant variation of $\tau_{D}$ with $\Delta$.

The intuitive understanding for these variations in the decay time scales can be understood from Eq.~\eqref{eq_tau_kitaev}. The inverse of the time scale associated with $\alpha$ (decaying) bath is proportional to the initial site occupancies, while for the $\beta$ (pumping) bath, it is proportional to the effective vacancy at any site. While a decaying bath will deplete a high-density state ($\mu \gg 1$) much more rapidly than a low-density state ($\mu \ll -1$) which will decay much more slowly, the situation reverses for $\alpha \rightarrow \beta$, $\mu \rightarrow -\mu$. Interestingly at $\alpha=\beta$, this decay rate is independent of $\mu$ (see Eq.~\eqref{eq_tau_kitaev}). Thus our results show that $\mu$ driven phase transitions in the Kitaev chain are decipherable in the early-time survival probability.

The results can also be interpreted from the analogy of a spin at some finite temperature $T_{\rm{eff}}$ and an external magnetic field $h$. Assuming $h \equiv (\beta-\alpha)$, $T_{\rm{eff}} \equiv (\alpha+\beta)$, Eq.~\eqref{eq_tau_kitaev} can be rewritten as
 \beq \label{eq_ts_map}
 \frac{1}{\tau_{D}}=  L\Big( -\frac{h}{2} {(m_{z})}_{0} + \frac{T_{\text{eff}}}{2} \Big),
 \eeq
thus leading to a thermal decay time scale and a field-driven time scale. Here, the effective magnetization of the initial state is given by
 \beq
 {(m_{z})}_{0}=\frac{1}{L} \sum_{j=1}^{L}\Tr (\rho (0) \sigma_{j}^{z})= \frac{1}{L} \sum_{j=1}^{L}\Tr (\rho (0) (2 c^{\dagger}_{j} c_{j}-1)),
 \eeq
which follows from the mapping of spin $\frac{1}{2}$ to spinless fermion, namely Jordan-Wigner transformation~\cite{lieb_1961}. With the increase of temperature, the spin is oriented more randomly leading to the decrease of survival probability with the timescale inversely proportional to $T_{\rm{eff}}$, irrespective of the presence of $h$. Now, if the external field $h$ is applied in the direction of the initial orientation of spin (i.e., $h {(m_{z})}_{0}>0$), $h$ favors the alignment of the spin in its initial direction, causing much slower decay of survival probability. On the other hand, if the external field $h$ is in the opposite direction of the orientation of the spin (i.e., $h {(m_{z})}_{0}<0$), $h$ flips the spin, which leads to faster decay of survival probability (see \Fig{fig:survprob_plot}(e)).

\begin{figure}
    \centering
\includegraphics[width=1\linewidth]{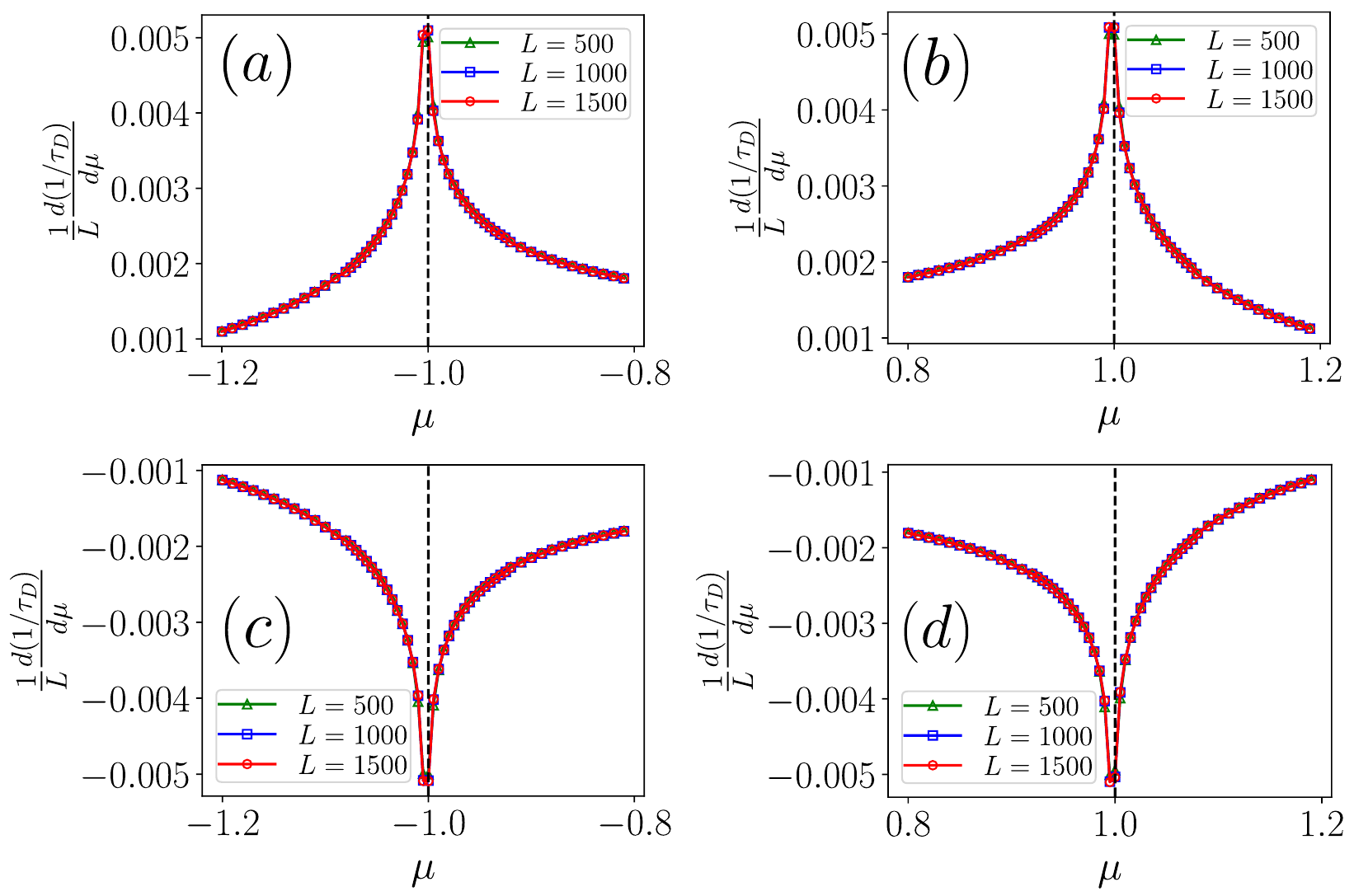}
\caption{\tb{$\frac{1}{L}\frac{d(1/\tau_{D})}{d \mu}$ (where the survival probability $F(t) \sim \exp(-t/\tau_{D})$)  as a function of chemical potential $\mu$ (a) near the critical point $\mu=-1$ for fermion-decaying bath ($\alpha=0.01$, $\beta=0$), (b) near the critical point $\mu=1$ for fermion-decaying bath ($\alpha=0.01$, $\beta=0$), (c) near the critical point $\mu=-1$ for fermion-pumping bath ($\alpha=0$, $\beta=0.01$), (d) near the critical point $\mu=1$ for fermion-pumping bath ($\alpha=0$, $\beta=0.01$). In all the plots, we have chosen $J=\Delta=1$.}}
    \label{fig:tsk}
\end{figure}

\tb{As we find that the timescale $\tau_{D}$ is proportional to $1/L$ (see Eq.~\eqref{eq_ts_map}) for all parameter regimes, the quantity $\frac{d(1/\tau_{D})}{d \mu}$ scales as $L$ for all parameters including the critical points. Thus, $\frac{1}{L}\frac{d(1/\tau_{D})}{d\mu}$ does not show divergence at $\mu=\pm 1$ with the increase of $L$ (see Fig.~\ref{fig:tsk}(a,b,c,d)) implying no singularities at the critical points in the thermodynamic limit. However, even in absence of any singular behavior as seen in conventional phase transitions of closed systems, the sharp changes of $\frac{1}{L}\frac{d(1/\tau_{D})}{d\mu}$ at $\mu=\pm 1$ enable us detecting the critical points $\mu=\pm 1$ in dissipative situations.} One consequence of having an  extensive number of sites coupled to the bath is that effective $\tau_D$ can be made vanishingly small ($\tau_{D}\propto \frac{1}{L}$) in the thermodynamic limit for the fermion decaying (pumping) bath for $\mu \gg 1$ ($\mu \ll -1$) (see Appendix~\ref{app_tau} for details).  In fact, depending on the number of sites the baths are coupled to, the survival probability can be tuned (see  Appendix~\ref{varybath} for details).

\begin{figure}
\centering
\includegraphics[width=1.0\linewidth]{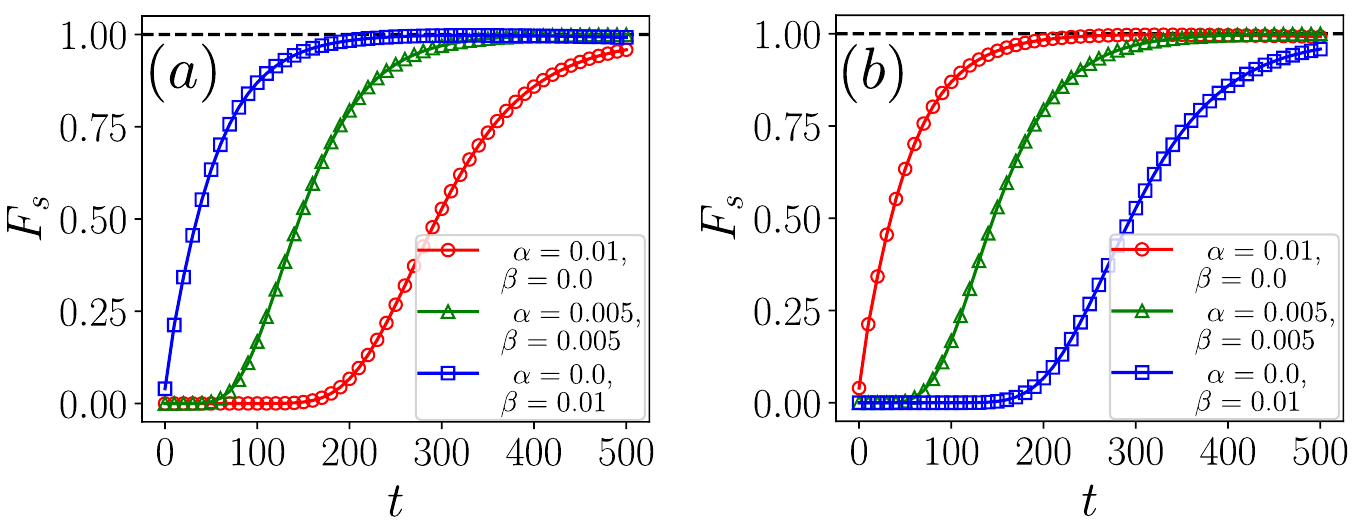}
\caption{Fidelity $F_{s}(t)$ between $\rho(t)$ and the steady state density matrix $\rho_{ss}$ (see Eq.~\eqref{eq_fss}) as a function of $t$ when (a) $\mu=1.5$, (b) $\mu=-1.5$. In both the plots, we have chosen $L=100$, $J=\Delta=1.0$.} 
\label{fig:ss_fidelity}
\end{figure}

While we have discussed fidelity in terms of overlap with the initial density matrix, it is also interesting to note how the density matrix approaches the final steady state. We therefore evaluate the overlap between $\rho(t)$ and the steady state density matrix $\rho_{ss}$ defined as
\begin{equation}\label{eq_fss}
F_{s}(t)=\Tr(\sqrt{\sqrt{\rho_{ss}} \rho(t) \sqrt{\rho_{ss}}}), 
\end{equation}
where both $\rho(t)$ and $\rho_{ss}$ are in general mixed states~\cite{jozsa_1994,nielsen_book}. Analyzing $F_{s}(t)$, we find that when $\mu \gg 1$, the approach to the steady state is faster for pumping bath ($\alpha=0$, $\beta \neq 0$) than for the decaying bath ($\alpha \neq 0$, $\beta=0$), as shown in Fig.~\ref{fig:ss_fidelity}(a). This can be explained as follows: pumping (decaying) bath leads to the steady state with higher (lower) site occupancies. When $\mu \gg 1$, the site occupancies in the ground state are close to $1$. Thus, the decaying bath takes much time to render the chain to its steady state compared to the pumping bath. The situation reverses when $\mu \ll -1$ (see Fig.~\ref{fig:ss_fidelity}(b)). When $\alpha=\beta$, the steady state is approached at the same rate for all $\mu$.

While we consider periodic boundary conditions for the analysis of survival probability, the behavior of survival probability remains similar in the corresponding open boundary conditions in the thermodynamic limit, as the jump operators are local in sites and identical for both periodic and open boundary conditions. However, we note that the topological properties in the dissipative systems can show a distinction between periodic and open boundary conditions due to edge modes in some fine-tuned situations like the existence of decoherence-free subspaces and disorder-induced localization ~\cite{bardyn2013topology,Carmele_PRB_2015,naren20}.

\section{Generalizations}
\label{generalBHZ}

\begin{figure*}
    \centering
\includegraphics[width=1\linewidth]{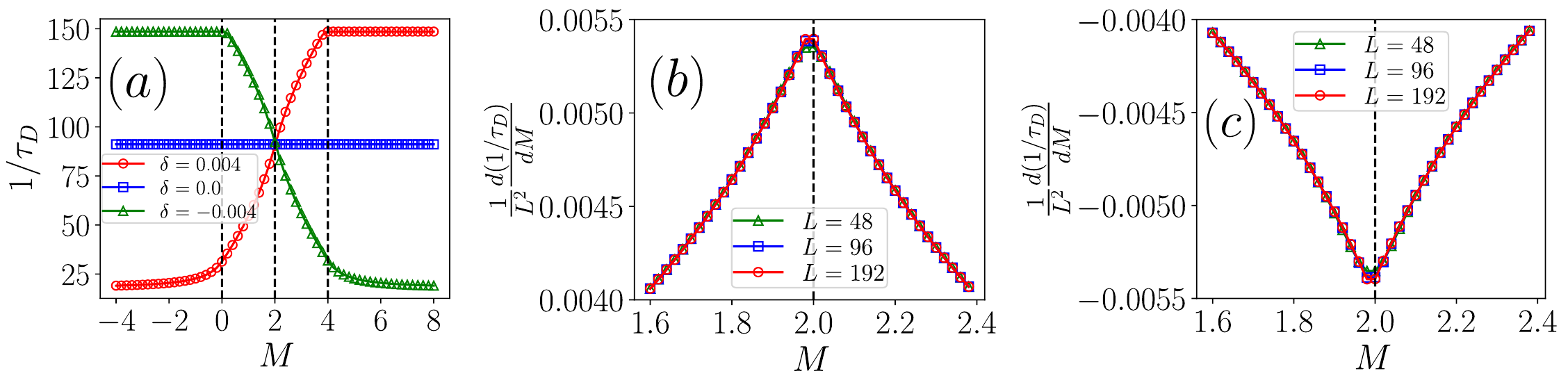}
\caption{(a) $1/\tau_{D}$ (where survival probability $F(t) \sim \exp(-t/\tau_{D})$) for BHZ model with size $L \times L$ as a function of $M$. In this plot, we have chosen $L=96$, $\alpha+\beta=0.01$. The dotted lines at $M=0,2,4$ are the critical lines. \tb{(b) $\frac{1}{L^{2}} \frac{d(1/\tau_{D})}{d M}$ as a function of $M$ near the critical point at $M=2$ when $\delta=0.004$, (c) $\frac{1}{L^{2}} \frac{d(1/\tau_{D})}{d M}$ as a function of $M$ near the critical point at $M=2$ when $\delta=-0.004$.}
}
    \label{fig:bhz}
\end{figure*}

In this section, we generalize the steady state and early time behavior of the dissipative systems to higher dimensions, particularly to the two-dimensional Bernevig-Hughes-Zhang (BHZ) model~\cite{bhz_2006} to show how such decipherable signatures of ground state phase transitions continue even in this system. BHZ model~\cite{bhz_2006} on a two-dimensional square lattice is described by the Hamiltonian
\begin{equation}\label{eq_hbhz}
H=(2-M) \sum_{j} {\bf{c}}_{j}^{\dagger} \sigma_{z} ~{\bf{c}}_{j} - \sum_{\langle jn \rangle} ({\bf{c}}_{j}^{\dagger} {\large{\eta}}_{jn}~{\bf{c}}_{n} + h.c.),
\end{equation}
where each unit cell contains $A$ and $B$ sites having staggered masses $(2-M)$ and $(-2+M)$, respectively. The hopping strengths between nearest neighbouring unit cells along $x$-axis and $y$-axis are ${\large{\eta}}_{jn}=\frac{1}{2} (\sigma_{z}+i \sigma_{x})$ and ${\large{\eta}}_{jn}=\frac{1}{2} (\sigma_{z}+i \sigma_{y})$, respectively and
${\bf{c}}_{j}=\begin{pmatrix}
c_{jA} & c_{jB}
\end{pmatrix}^{\top}$. Time-reversal symmetry and sublattice symmetry are absent in the BHZ model and thus it belongs to the symmetry class $D$ in the tenfold symmetry classification~\cite{altland_prb_97}. Quantum critical points (QCPs) of the BHZ model are situated at $M=0,2,4$ where the bulk energy-gap closes.
For the BHZ model in the presence of baths, we consider the dissipator of the form
\begin{multline}
{\mathcal{D}} = \alpha \sum_{n} \left({\mathcal{D}} [{\mathcal{L}}_{n}=c_{nA}] + {\mathcal{D}} [{\mathcal{L}}_{n}=c_{nB}]\right) \\
+ \beta \sum_{n} \left({\mathcal{D}} [{\mathcal{L}}_{n}=c^{\dagger}_{nA}] + {\mathcal{D}} [{\mathcal{L}}_{n}=c_{nB}^{\dagger}]\right),
\end{multline}
where $n$ runs over all the unit cells. Similar to the Kitaev chain, $\alpha$ and $\beta$ are the coupling strengths of the decaying and the pumping bath, respectively.

\subsection{Steady state}
For the periodic boundary condition of the BHZ model in the presence of baths, the Lindblad equation for the momentum ${\bf{k}}=(k_{x},k_{y})$ (where $k_{x} \in [-\pi,\pi]$, $k_{y} \in [-\pi,\pi]$) can be written as
\begin{multline}
\frac{d \rho^{{\bf{k}}}}{dt}= -i [H_{\bf{k}}, \rho^{{\bf{k}}}] + \alpha \left( c_{{\bf{k}}A} \rho^{{\bf{k}}} c_{{\bf{k}}A}^{\dagger} - \frac{1}{2} \{ c_{{\bf{k}}A}^{\dagger} c_{{\bf{k}}A} , \rho^{{\bf{k}}} \} \right) \\
+ \alpha \left( c_{{\bf{k}}B} \rho^{{\bf{k}}} c_{{\bf{k}}B}^{\dagger} - \frac{1}{2} \{ c_{{\bf{k}}B}^{\dagger} c_{{\bf{k}}B} , \rho^{{\bf{k}}} \} \right) \\
+ \beta \left( c_{{\bf{k}}A}^{\dagger} \rho^{{\bf{k}}} c_{{\bf{k}}A} - \frac{1}{2} \{ c_{{\bf{k}}A} c_{{\bf{k}}A}^{\dagger} , \rho^{{\bf{k}}} \} \right) \\
+ \beta \left( c_{{\bf{k}}B}^{\dagger} \rho^{{\bf{k}}} c_{{\bf{k}}B} - \frac{1}{2} \{ c_{{\bf{k}}B} c_{{\bf{k}}B}^{\dagger} , \rho^{{\bf{k}}} \} \right),
\end{multline}
where the Hamiltonian $H_{\bf{k}}$ is given by
\beq
\begin{split}\notag
H_{\bf{k}}&=\left( 2-M-\cos(k_{x})- \cos(k_{y}) \right) \left( c_{{\bf{k}}A}^{\dagger} c_{{\bf{k}}A} - c_{{\bf{k}}B}^{\dagger} c_{{\bf{k}}B} \right)\\
&+\sin(k_{x}) \left( c_{{\bf{k}}A}^{\dagger} c_{{\bf{k}}B} + c_{{\bf{k}}B}^{\dagger} c_{{\bf{k}}A} \right)\\
&- i \sin(k_{y}) \left( c_{{\bf{k}}A}^{\dagger} c_{{\bf{k}}B} - c_{{\bf{k}}B}^{\dagger} c_{{\bf{k}}A} \right).
\end{split}
\eeq
The steady state density matrix $\rho^{\bf {k}}_{ss}$ of the BHZ model is then obtained by $\frac{d \rho_{ss}^{\bf{k}}}{dt}=0$, where $\rho_{ss}^{\bf{k}}$ can be represented by $4 \times 4$ matrix in the basis $| 0,0 \rangle$, $| {\bf{k}} A, 0 \rangle$, $| 0, {\bf{k}} B \rangle$ and $| {\bf{k}} A, {\bf{k}} B \rangle$. Depending on different values of $\alpha$ and $\beta$, the steady states can be either a pure state or a mixed state. We find that the nature of the steady state for the BHZ model exactly follows the behavior as shown in Table~\ref{tab:ssans} for $\Delta=0$ in the Kitaev chain. Similar to the steady states for the dissipative Kitaev chain with $\Delta=0$, even here the presence of only $\alpha$ or $\beta$ bath leads to a pure state. Again when $\alpha=\beta$, the steady state is an infinite temperature thermal state. Interestingly when $\alpha \neq \beta \neq 0$, the steady state is a finite temperature mixed state as shown in Table~\ref{tab:ssans}. The results therefore qualify the insights uncovered in the Kitaev model of the previous section with $\Delta=0$. In the next section, we briefly discuss the survival probabilities of the BHZ model.  

\subsection{Survival probability}
Similar to the Kitaev chain, the survival probability  $F(t)$ for the BHZ model in the presence of bath decays exponentially with time at early times, i.e., $F(t) \sim \exp(-t/\tau_{D})$. Here again, $\tau_D$ can be related to fermionic densities of the initial state:
\begin{equation}
\frac{1}{\tau_{D}}=   L^{2} \alpha \left( {\langle  n_{A} \rangle}_{0} + {\langle  n_{B} \rangle}_{0} \right) +  L^{2} \beta \left( 2- {\langle n_{A} \rangle}_{0} - {\langle n_{B} \rangle}_{0} \right),
\end{equation}
where ${\langle n_{A} \rangle}_{0}$ and ${\langle n_{B} \rangle}_{0}$ are the expectation values of the fermionic densities $n_{A}=\frac{1}{L^{2}} \sum_{j} c_{jA}^{\dagger} c_{jA}$ and $n_{B}=\frac{1}{L^{2}} \sum_{j} c_{jB}^{\dagger} c_{jB}$ in sublattices $A$ and $B$ respectively in the initial state and $L^{2}$ is the number of unit cells. Given the quantum phase transitions in the BHZ model are engineered by the parameter $M$, $\tau_D$ in general, is independent of $M$ (as ${\langle  n_{A} \rangle}_{0} + {\langle  n_{B} \rangle}_{0}=1$ for all $M$). However, a sublattice selective bath couplings can lead to different values of $\tau_D$ depending on the initial state. Introducing an asymmetric parameter in the bath couplings $\delta$ between the two sublattices such that the coupling strengths for decaying baths on the sublattices A and B are $(\alpha+\delta)$ and $(\alpha-\delta)$, while for the pumping baths the coupling strengths are $(\beta-\delta)$ and $(\beta+\delta)$ respectively, one obtains
\begin{multline}\label{eq_asymtau}
\frac{1}{\tau_{D}}=   L^{2} \alpha \left( {\langle  n_{A} \rangle}_{0} + {\langle  n_{B} \rangle}_{0} \right) + 2L^{2}\delta \left( {\langle  n_{A} \rangle}_{0} - {\langle  n_{B} \rangle}_{0} \right) \\
+  L^{2} \beta \left( 2- {\langle n_{A} \rangle}_{0} - {\langle n_{B} \rangle}_{0} \right).
\end{multline}
Here when $\delta >0$, one finds the $1/\tau_D$ behavior can capture the critical point at $M=4$ and the critical point $M=0$ is captured with $\delta<0$. From the crossing of $1/\tau_{D}$ for $\delta>0$ and $\delta<0$, the critical point at $M=2$ is also detected. Clearly when $\delta=0$ the phase transitions cannot be captured (see Fig.~\ref{fig:bhz}(a)).

\tb{It is also important to note here that $1/\tau_{D}$ scales as $L^{2}$ (see Eq.~\eqref{eq_asymtau}) and thus the quantity $\frac{d(1/\tau_{D})}{d M}$ is also proportional to $L^{2}$ for all values of $M$ including the critical points. This leads to the non-diverging behavior of $\frac{1}{L^2} \frac{d(1/\tau_{D})}{d M}$ at the critical points with 
increasing $L$ for both $\delta>0$ and $\delta<0$ (see Fig.~\ref{fig:bhz}(b) and Fig.~\ref{fig:bhz}(c)). However, even in the absence of any singular behavior at the critical points in the thermodynamic limit, it is interesting to note that sharp changes in the inverse decay time scale ($1/\tau_D$) carry signatures of underlying quantum phase transitions of the ground state. }


\section{Outlook}
\label{sec:outlook}

The study of quantum many-body systems which are coupled to external baths is of interest both from the theoretical questions it poses, and the experimental avenues such as cold atoms, Floquet systems, and in material settings. The state of such quantum systems is typically understood under a Lindbladian evolution of the density matrix. While such a steady-state density matrix is generally mixed, it is often not clear what physics it entails and whether it can carry any signature of the ground state phases of the system. In this work, we have studied the steady state of a paradigmatic toy model, the Kitaev chain under the combined effects of both particle pulling and pumping baths. The problem can be exactly solved within methods such as third quantization given the system retains the structure of being a quadratic Lindbladian. However, our work uncovers an effective mapping of the system to a classical spin in an effective temperature and a magnetic field, thus making the nature of decoherence processes transparent. We then show that the presence of a superconducting pairing term leads to non-trivial correlations which lead to decipherable signatures of the ground state critical point even in the steady state. This is the central result of our work. The critical properties of the phase transition lines can also be captured quite sharply (despite not being exact singularities like in unitary systems), with the inclusion of some broadening effects due to effective decoherence scales. We also show that the interplay of the decay and pumping parameters can modulate the steady state properties and also change its entanglement character. 
\tb{Particularly, the dissipation in the system due to the coupling with the bath leads to an additional volume-law contribution of entanglement entropy, which is in contrast to the area-law contribution and logarithmic correction of entanglement entropy in the ground state of gapped and gapless closed systems. Further, unlike the power-law decay of two-point correlation and a diverging correlation length at the gapless critical points of a closed system, the dissipation in an open system leads to an exponentially decaying two-point correlations in the steady state along with a non-diverging correlation length even at the critical points.}

While discussed in the specific context of the Kitaev model, our results show that these are rather general features available to free fermionic systems, even in particle number-conserving systems. We verify this in the BHZ model which is a two-dimensional Chern insulator. While the steady state physics itself points out interesting signatures of the ground state phase diagram, we also show that the early time fidelity can parallelly signal some phase transitions. This has to do with the decay time scales associated with the observables and the initial density matrix. This analysis shows that both the chemical potential driven phase transitions in the case of the Kitaev model, as well as the topological phase transitions in the BHZ model can be deciphered for appropriately chosen bath couplings. Taken together our work emphasizes that even under Lindbladian evolution, signatures of ground state phases get realized in both the early time dynamics and in steady states. 

\tb{While we explore the steady states and in early-time relaxation dynamics in this paper, it would be interesting to see if there is any signature of quantum criticality in long-time relaxation dynamics.} Also, as the presence of two bath parameters $\alpha$  and $\beta$ in our system could be mapped to an effective temperature $\sim (\alpha +\beta)$ and field $(\beta-\alpha)$, it will be an interesting direction to see the behavior of such Lindbladian systems under a more general number of bath parameters and larger Hamiltonian parameter space. This might lead to the realization of systems with multiple temperature scales (decoherence scales governed by bath parameters) not realizable in standard finite temperature  equilibrium  systems. \tb{Further, the generalization of the results of this paper to other many-body systems would also be an important future direction. Specifically, whether the steady state of other dissipative many-body systems can still contain the signatures of initial state is a matter of further investigation. However, we anticipate that when a dissipator in the non-interacting system leads to non-thermal steady state, the observables evaluated in the steady state will in general carry the signatures of the initial states.}
The inclusion of other types of dissipators (e.g., non-Markovian dissipators) and a systematic study of the corresponding departure of the phase space trajectories from the Lindbladian prediction is also an exciting future prospect. 

\begin{acknowledgements}
We honor and dedicate this work to the loving memory of Prof.~Amit Dutta who seeded many of our interests in the questions addressed in this work. We acknowledge fruitful discussions with Diptarka Das, Suraj Hegde, Harish Adsule, Saikat Ghosh, Parveen Kumar, Adolfo del Campo and Arnab Das.  R. Joshi acknowledges financial support from Indian Institute of Technology Kanpur as well as from the Department of Physics, University of Illinois Urbana-Champaign. SM acknowledges financial support from
PMRF fellowship, MHRD, India. S. Bandyopadhyay acknowledges Boston University and PMRF, MHRD, India for financial support. S. Bhattacharjee acknowledges support from: European Research Council AdG NOQIA; MCIN/AEI (PGC2018-0910.13039/501100011033, CEX2019-000910-S/10.13039/501100011033, Plan National FIDEUA PID2019-106901GB-I00, Plan National STAMEENA PID2022-139099NB-I00 project funded by MCIN/AEI/10.13039/501100011033 and by the “European Union NextGenerationEU/PRTR" (PRTR-C17.I1), FPI); QUANTERA MAQS PCI2019-111828-2); QUANTERA DYNAMITE PCI2022-132919 (QuantERA II Programme co-funded by European Union’s Horizon 2020 program under Grant Agreement No 101017733), Ministry of Economic Affairs and Digital Transformation of the Spanish Government through the QUANTUM ENIA project call – Quantum Spain project, and by the European Union through the Recovery, Transformation, and Resilience Plan – NextGenerationEU within the framework of the Digital Spain 2026 Agenda; Fundació Cellex; Fundació Mir-Puig; Generalitat de Catalunya (European Social Fund FEDER and CERCA program, AGAUR Grant No. 2021 SGR 01452, QuantumCAT \ U16-011424, co-funded by ERDF Operational Program of Catalonia 2014-2020); Barcelona Supercomputing Center MareNostrum (FI-2023-1-0013); Funded by the European Union. Views and opinions expressed are, however, those of the author(s) only and do not necessarily reflect those of the European Union, European Commission, European Climate, Infrastructure and Environment Executive Agency (CINEA), or any other granting authority. Neither the European Union nor any granting authority can be held responsible for them (EU Quantum Flagship (PASQuanS2.1, 101113690); EU Horizon 2020 FET-OPEN OPTOlogic (Grant No 899794)); EU Horizon Europe Program (This project has received funding from the European Union’s Horizon Europe research and innovation program under grant agreement No 101080086 NeQSTGrant Agreement 101080086 — NeQST)‌, ICFO Internal “QuantumGaudi” project; European Union’s Horizon 2020 program under the Marie Sklodowska-Curie grant agreement No 847648; “La Caixa” Junior Leaders fellowships, La Caixa” Foundation (ID 100010434): CF/BQ/PR23/11980043. AA acknowledges support from IITK Initiation Grant (IITK/PHY/2022010). Numerical calculations were performed on the workstation {\it Wigner} at IITK.
\end{acknowledgements}

\appendix

\section{Steady state density matrix in momentum space for Kitaev chain}\label{app_sskt}
 Defining $\Delta^{\prime} = \Delta\sin(k)$,  $\epsilon^{\prime} = \mu + J\cos(k)$, the non-zero elements of the steady state density matrix ($\rho^{k}_{ss}$) are:
   \begin{align}
        \rho_{k}^{11}(\infty) &= \frac{{(\Delta^\prime)}^2{(\alpha+ \beta)}^2 + \alpha^2 {(\alpha + \beta)}^2 + 4\alpha^{2}{(\epsilon^{\prime})}^{2}}{4{(\Delta^\prime)}^2{(\alpha + \beta)}^2 + {(\alpha + \beta)}^4 + 4{(\epsilon^{\prime})}^{2}{(\alpha + \beta)}^2},\\
        \rho_k^{22}(\infty) &= \frac{{(\Delta^\prime)}^2{(\alpha+ \beta)}^2 + \alpha\beta {(\alpha + \beta)}^2 + 4\alpha\beta{(\epsilon^{\prime})}^{2}}{4{(\Delta^\prime)}^2{(\alpha + \beta)}^2 + {(\alpha + \beta)}^4 + 4{(\epsilon^{\prime})}^{2}{(\alpha + \beta)}^2},\\
        \rho_k^{33}(\infty) &= \frac{{(\Delta^\prime)}^2{(\alpha+ \beta)}^2 + \alpha\beta {(\alpha + \beta)}^2 + 4\alpha\beta{(\epsilon^{\prime})}^{2}}{4{(\Delta^\prime)}^2{(\alpha + \beta)}^2 + {(\alpha + \beta)}^4 + 4{(\epsilon^{\prime})}^{2}{(\alpha + \beta)}^2},\\
        \rho_k^{44}(\infty) &= \frac{{(\Delta^\prime)}^2{(\alpha+ \beta)}^2 + \beta^2 {(\alpha + \beta)}^2 + 4\beta^2{(\epsilon^{\prime})}^{2}}{4{(\Delta^\prime)}^2{(\alpha + \beta)}^2 + {(\alpha + \beta)}^4 + 4{(\epsilon^{\prime})}^{2}{(\alpha + \beta)}^2},\\
    \Re\rho_k^{14}(\infty) &= \frac{{(\Delta^\prime)} (\beta- \alpha)}{4{(\Delta^\prime)}^2 + {(\alpha + \beta)}^2 + 4{(\epsilon^{\prime})}^{2}},\\
    \Im\rho_k^{14}(\infty) &= \frac{2{(\Delta^\prime)}{(\epsilon^{\prime})}(\alpha- \beta)}{4{(\Delta^\prime)}^2(\alpha + \beta) + {(\alpha + \beta)}^3 + 4{(\epsilon^{\prime})}^{2}(\alpha+\beta) }.
\end{align}

\section{Analytical calculation of time-scale $\tau_{D}$ for Kitaev chain}\label{app_tau}
As the initial state (at $t=0$) is a pure state (i.e., $\rho^{2}(0)=\rho(0)$), the time evolution of survival probability $F(t)$ defined in Eq.~\eqref{eq_sp} at small time ($t \to 0$) is given by
\begin{multline}\label{eq_dfdtf}
\frac{d F(t)}{dt} \approx \alpha \sum_{n=1}^{L} \left(\Tr \left(c_{n} \rho(0)\right) \Tr \left(c^{\dagger}_{n} \rho(0)\right) -\Tr\left(c^{\dagger}_{n} c_{n} \rho(0)\right)\right) \\
 + \beta \sum_{n=1}^{L} \left(\Tr\left(c^{\dagger}_{n} \rho(0)\right) \Tr \left( c_{n} \rho(0)\right) -\Tr \left(c_{n} c^{\dagger}_{n} \rho(0)\right)\right).
\end{multline}
As $\Tr \left(c_{n} \rho(0)\right) = {\langle c_{n} \rangle}_{0}=0$ and $\Tr \left(c^{\dagger}_{n} \rho(0)\right) = {\langle c^{\dagger}_{n} \rangle}_{0}=0$,  Eq.~\eqref{eq_dfdtf} reduces to
\begin{equation}
\frac{d F(t)}{dt} \approx -\frac{1}{\tau_{D}}.
\end{equation}
Here, the time-scale $\tau_{D}$ is determined by
\begin{align}
\frac{1}{\tau_{D}}= \sum_{n=1}^{L} \left(\alpha \Tr (c^{\dagger}_{n} c_{n} \rho(0)) + \beta \Tr ( c_{n} c^{\dagger}_{n} \rho(0))\right) \nonumber
\\ = \sum_{n=1}^{L} \left(\alpha {\langle c^{\dagger}_{n} c_{n} \rangle}_{0} + \beta \left(1-{\langle c^{\dagger}_{n} c_{n} \rangle}_{0}\right)\right), \nonumber
\end{align}
where the expectation value ${\langle c^{\dagger}_{n} c_{n} \rangle}_{0}$ is calculated in the initial ground state of the Kitaev chain (at time $t=0$ in the absence of a bath).
When $\mu \gg 1$, ${\langle c^{\dagger}_{n} c_{n} \rangle}_{0} =1$ and when $\mu \ll -1$, ${\langle c^{\dagger}_{n} c_{n} \rangle}_{0} =0$ (for all $n$). Thus, For fermion-decaying bath ($\alpha \neq 0$, $\beta=0$),
\begin{equation}
    1/\tau_{D} \approx \begin{cases}
		 \alpha L , & \text{for $\mu \gg 1$},  \\
		0 , & \text{for $\mu \ll -1$},
	\end{cases}
\end{equation}
and for fermion-pumping bath ($\alpha=0$, $\beta \neq 0$),
\begin{equation}
    1/\tau_{D} \approx \begin{cases}
		0 , & \text{for $\mu \gg 1$},  \\
		 \beta L , & \text{for $\mu \ll -1$}.
	\end{cases}
\end{equation}
Therefore, the survival probability $F(t)$ decays faster (slower) in the presence of a fermion-decaying bath than in the presence of a fermion-pumping bath when $\mu>1$ ($\mu<-1$). Further, for $\alpha=\beta$, we find $1/\tau_{D}=\alpha L$, which confirms that $\tau_{D}$ is independent of $\mu$ when $\alpha=\beta$.

\section{Varying the number of sites coupled to baths}\label{varybath}

\begin{figure}
    \centering
\includegraphics[width=0.8\linewidth]{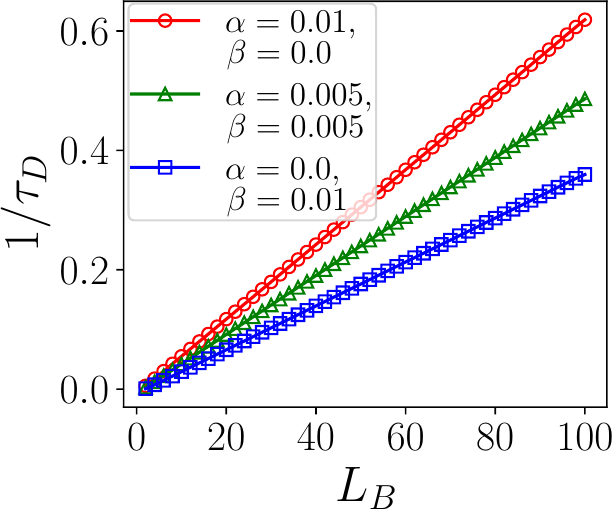}
     \caption{$1/\tau_{D}$ (where survival probability $F(t) \sim \exp(-t/\tau_{D})$) as a function of number of sites $L_{B}$ coupled to the bath. In this plot, we have chosen open boundary condition of the Kitaev chain with $J=\Delta=1.0$, $\mu=0.5$, $L=100$.}
    \label{fig:surv_prob_obc}
\end{figure}

Let us consider the Kitaev chain of $L$ sites with open boundary conditions and $L_{B}$ number of sites of the chain coupled to the bath. Thus, the dissipator can be written as
\begin{equation}
{\mathcal{D}}= \alpha \sum_{n=1}^{L_{B}} {\mathcal{D}} [{\mathcal{L}}_{n}=c_{n}] + \beta \sum_{n=1}^{L_{B}}{\mathcal{D}} [{\mathcal{L}}_{n}=c_{n}^{\dagger}],
\end{equation}
where ${\mathcal{D}}[{\mathcal{L}}_{n}]= \left( {\mathcal{L}}_{n} \rho {\mathcal{L}}_{n}^{\dagger}-\frac{1}{2} \{ {\mathcal{L}}_{n}^{\dagger} {\mathcal{L}}_{n} , \rho \} \right)$. Similar to the situation with periodic boundary conditions, here also the survival probability $F(t)$ decays exponentially with time (i.e. $F(t) \sim \exp(-t/\tau_{D})$). The behaviors of $\tau_{D}$ for different choices of $\alpha$ and $\beta$ are also similar to that in the periodic boundary condition. Further, we find that $1/\tau_{D}$ increases linearly with the number of sites $L_{B}$ coupled to bath (see Fig.~\ref{fig:surv_prob_obc}), implying faster decay of survival probability with the increase of sites coupled to the bath.

\bibliography{ref_dissipative}

\end{document}